# Tailoring Scaffolding to Diagnostic Strategies: Theory-Informed LLM-Based Agents

FATMA BETÜL GÜRES, EPFL, ETH Zürich, Switzerland
TANYA NAZARETSKY, EPFL, Switzerland
TANJA KÄSER, EPFL, Switzerland

**ABSTRACT**: Learning analytics systems increasingly integrate large language models (LLMs) to provide adaptive scaffolding in complex learning environments, yet personalization is often driven by global instructional choices rather than principled alignment with learning theory, limiting effectiveness and pedagogical grounding. In prior work, we examined how structuring and problematizing scaffolding approaches can be instantiated through LLM agents in a scenario-based learning environment for diagnostic reasoning. While both approaches supported learning, we observed systematic differences in learner interaction patterns and clear tendencies indicating that different diagnostic strategies benefited from distinct forms of scaffolding. Building on these findings, we propose a theory-informed scaffolding design grounded in the Knowledge Learning Instruction (KLI) framework, as different diagnostic strategies target different types of knowledge and require different instructional mechanisms. We use KLI to guide the alignment between strategy demands and scaffolding approaches and introduce a KLI-informed hybrid LLM agent that adapts its pedagogical support according to the diagnostic strategy being practiced, rather than applying a single global scaffolding approach. We hypothesize that this design could enable better learning gains. The study is scheduled for January. In the case of acceptance, preliminary results will be presented at the workshop.

## 1 INTRODUCTION

Learning analytics systems increasingly integrate large language models (LLMs) to provide adaptive scaffolding in complex learning environments. While these systems promise personalization at scale, many existing approaches rely on global instructional choices or surface-level behavioral signals, rather than principled alignment with learning theories. This limits both the interpretability of adaptive decisions and their pedagogical grounding. In our prior work (Authors, 2026), we examined how two theory-driven scaffolding approaches, structuring and problematizing (Reiser, 2004), can be instantiated through an LLM-based mentor in a scenario-based learning environment to enhance learners' diagnostic reasoning by guiding them to apply diagnostic strategies. Although both approaches supported learning, we observed systematic differences in learner interaction patterns and clear tendencies suggesting that different diagnostic strategies benefited from different forms of scaffolding. These findings indicate that applying a single, fixed scaffolding approach may not adequately support the diversity of reasoning processes involved in diagnostic tasks. Building on this insight, this workshop paper argues for a shift from globally defined scaffolding strategies toward theory-informed, strategy-aligned support. Specifically, we propose the Knowledge-Learning-Instruction (KLI) framework (Koedinger et al, 2012) as a lens for understanding how different diagnostic strategies target distinct types of knowledge and, therefore, require particular instructional mechanisms. From this perspective, structuring and problematizing should be seen as complementary forms of support, selectively deployed based on the learner's current diagnostic strategy rather than treated as competing alternatives. We use this theoretical framing to propose a KLI-informed hybrid LLM agent that adapts its pedagogical support according to the diagnostic strategy being practiced. Rather than enforcing a single scaffolding style, the agent dynamically adjusts between structuring-oriented and problematizing-oriented support in response to learners' evolving knowledge levels and strategies to master. We hypothesize that this approach can enable more effective and interpretable adaptive scaffolding by aligning instructional support more closely with the cognitive demands of a particular type of diagnostic reasoning. The goal of this contribution is to outline a theory-driven extension of our prior work and to invite discussion on how learning theories such as KLI can guide the design of adaptive ol, pedagogically grounded LLM-based agents in learning analytics systems. To structure this follow-up work, we formulate the following research questions: **RQ1**. How do different LLM-based scaffolding approaches (structuring, problematizing, and a KLI-informed hybrid) shape learners' interaction patterns and engagement during diagnostic reasoning tasks? **RQ2**. To what extent does strategy-aligned, KLI-informed scaffolding influence learning outcomes compared to globally applied scaffolding approaches?

## 2 METHODOLOGY

**Study environment: PharmaSim Switch.** PharmaSim Switch (Authors, 2026) is a scenario-based learning application for pharmacy assistant training in which learners practice diagnostic reasoning through natural-language interactions with simulated clients and personalized interactive feedback from an LLM-based pharmacist mentor. Learners work through realistic client cases that require systematic information gathering and interpretation.

**Diagnostic Strategies in PharmaSim Switch.** PharmaSim Switch is designed to teach three diagnostic strategies. The **Checklist Strategy** focuses on ensuring comprehensive coverage of relevant symptoms, contextual factors, and red flags, emphasizing procedural completeness rather than interpretation. This strategy aligns naturally with the use of *structuring* scaffolds, which support task organization, sequencing, and coverage. **The Interpersonal Relationships Strategy** involves identifying the diagnostically relevant actor in a given scenario, such as the patient or a caregiver, and directing questions accordingly. Because this strategy requires recognizing relational structure before effective information gathering can occur, it is supported through an initial *problematizing* scaffold that prompts reflection on whom to question and why, followed by *structuring* scaffolds to guide systematic inquiry. **The Possible Causes Strategy** focuses on generating, comparing, and justifying potential explanations for the client's condition, requiring articulation of causal reasoning and assessment of plausibility. This strategy aligns primarily with *problematizing* scaffolds, which encourage explanation, comparison, and justification.

**Scaffolding Approaches Across Strategies.** Scaffolding in PharmaSim Switch is delivered through an LLM-based pharmacist agent and is adapted to the diagnostic strategy being practiced. In the **structuring condition**, the agent consistently applies structuring scaffolds across all strategies, emphasizing task organization, sequencing, and constraint of the problem space. In the **problematizing condition**, the agent emphasizes epistemic challenge throughout the interaction by prompting learners to explain, justify, and reflect on their reasoning. The **hybrid condition** implements strategy-aligned scaffolding by providing different forms of support depending on the cognitive role of each strategy. Checklist-based strategies receive *structuring* support to ensure systematic coverage, interpersonal reasoning is supported through *problematizing,* followed by *structuring* guidance, and possible causes reasoning is supported through *problematizing* scaffolding that promotes articulation and justification. This design operationalizes strategy-aligned scaffolding rather than applying a single global approach.

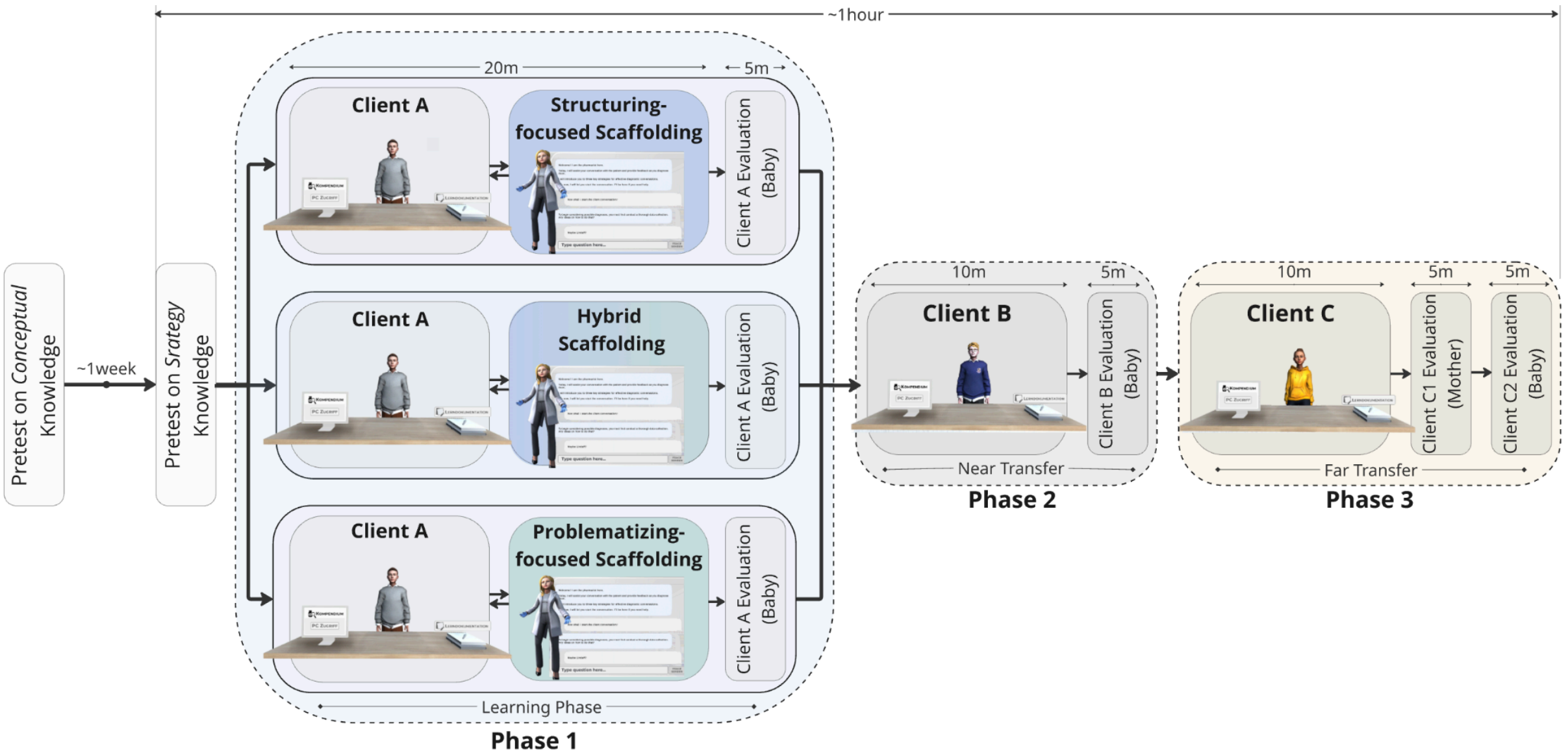

**Figure 1: Experimental design. Pretests are followed by a learning phase with Structuring-focused, Problematizing-focused, or Hybrid scaffolding (Phase 1) and two transfer phases (Phases 2–3).**

As illustrated in Figure 1, the study adopts a between-groups experimental design in which learners first complete pretests, followed by a learning phase featuring Structuring-focused, Problematizing-focused, or Hybrid scaffolding, and finally two subsequent transfer phases to measure students' learning in a condition without scaffolding. This design enables a systematic comparison of learners' ability to apply diagnostic strategies across novel contexts during transfer. By combining this experimental structure with a theory-informed scaffolding design grounded in the KLI framework, the study aims to inform the design of adaptive LLM-based agents that are pedagogically grounded.

## 3. Expected contribution

This research provides a concrete demonstration of how the KLI Framework can be operationalized to align the teaching of diagnostic strategies with scaffolding design and personalization within an LLM-based learning analytics system. We hypothesize that a KLI-informed design of the PharmaSim mentor agent may lead to improved learning outcomes and more effective diagnostic reasoning processes. A study involving 173 pharmacy technician apprentices was conducted in January 2026, and the results are currently under analysis and will be reported in future work.